\documentclass[aps, prd, amsmath, floats,floatfix, superscriptaddress, nofootinbib, showpacs]{revtex4}

\usepackage{amssymb}
\usepackage{amsmath}
\usepackage{verbatim}
\usepackage{mathrsfs}
\usepackage{amsfonts}
\usepackage{latexsym}
\usepackage{epsfig}
\usepackage{epstopdf}
\usepackage{color}
\usepackage{graphicx,subfigure}
\usepackage{units}
\usepackage{overpic}

\newcommand{\re}{\mbox{Re}}

\begin{document}


\definecolor{orange}{rgb}{0.9,0.45,0} 

\newcommand{\argelia}[1]{\textcolor{red}{{\bf Argelia: #1}}}
\newcommand{\dario}[1]{\textcolor{red}{{\bf Dario: #1}}}
\newcommand{\juanc}[1]{\textcolor{green}{{\bf JC: #1}}}
\newcommand{\juan}[1]{\textcolor{cyan}{{\bf Juan B: #1}}}
\newcommand{\alberto}[1]{\textcolor{blue}{{\bf Alberto: #1}}}
\newcommand{\miguela}[1]{\textcolor{red}{{\bf Miguel: #1}}}
\newcommand{\mm}[1]{\textcolor{orange}{{\bf MM: #1}}}
\newcommand{\OS}[1]{\textcolor{blue}{{\bf Olivier: #1}}}

\long\def\symbolfootnote[#1]#2{\begingroup%
\def\thefootnote{\fnsymbol{footnote}}\footnote[#1]{#2}\endgroup}


\title{$\ell$-Boson stars}

\author{Miguel Alcubierre}
\affiliation{Instituto de Ciencias Nucleares, Universidad Nacional Aut\'onoma de M\'exico,
Circuito Exterior C.U., A.P. 70-543, M\'exico D.F. 04510, M\'exico}

\author{Juan Barranco}
\affiliation{Departamento de F\'isica, Divisi\'on de Ciencias e Ingenier\'ias,
Campus Le\'on, Universidad de Guanajuato, Le\'on 37150, M\'exico}

\author{Argelia Bernal}
\affiliation{Departamento de F\'isica, Divisi\'on de Ciencias e Ingenier\'ias,
Campus Le\'on, Universidad de Guanajuato, Le\'on 37150, M\'exico}

\author{Juan Carlos Degollado}
\affiliation{Instituto de Ciencias F\'isicas, Universidad Nacional Aut\'onoma de M\'exico,
Apdo. Postal 48-3, 62251, Cuernavaca, Morelos, M\'exico}

\author{Alberto Diez-Tejedor}
\affiliation{Departamento de F\'isica, Divisi\'on de Ciencias e Ingenier\'ias,
Campus Le\'on, Universidad de Guanajuato, Le\'on 37150, M\'exico}

\author{Miguel Megevand}
\affiliation{Instituto de F\'isica Enrique Gaviola, CONICET. Ciudad Universitaria, 5000 C\'ordoba, Argentina}

\author{Dar\'io N\'u\~nez}
\affiliation{Instituto de Ciencias Nucleares, Universidad Nacional Aut\'onoma de M\'exico,
Circuito Exterior C.U., A.P. 70-543, M\'exico D.F. 04510, M\'exico}

\author{Olivier Sarbach}
\affiliation{Instituto de F\'isica y Matem\'aticas, Universidad Michoacana de San Nicol\'as de Hidalgo,
Edificio C-3, Ciudad Universitaria, 58040 Morelia, Michoac\'an, M\'exico}


\date{\today}


\begin{abstract}  
We present new, fully nonlinear numerical solutions to the static,
spherically symmetric Einstein-Klein-Gordon system for a collection of
an arbitrary odd number $N$ of complex scalar fields with an internal
$U(N)$ symmetry and no self-interactions. These solutions, which we
dub $\ell$-boson stars, are parametrized by an angular momentum number
$\ell=(N-1)/2$, an excitation number $n$, and a continuous parameter
representing the amplitude of the fields. They are regular at every
point and possess a finite total mass. For $\ell = 0$ the standard
spherically symmetric boson stars are recovered. We determine their
generalizations for $\ell > 0$, and show that they give rise to a
large class of new static configurations which might have a much
larger compactness ratio than $\ell=0$ stars.
\end{abstract}




\maketitle


\section{Introduction}
\label{sec:introduction}

Boson stars composed of massive scalar fields are among the most
promising exotic objects that may populate the universe. They were
originally proposed by Kaup~\cite{Kaup68} and Ruffini and
Bonazzola~\cite{Ruffini:1969qy} in the late sixties, and explored in
more detail during the subsequent decades in e.g.
Refs.~\cite{Colpi:1986ye,Friedberg87,Gleiser:1988rq,Lee:1988av,Seidel:1990,Guzman:2004wj}
(see
also~\cite{Jetzer:1991jr,Lee:1991ax,Schunck:2003kk,Liebling:2012fv}
for reviews). Even though they remain hypothetical, boson stars are
frequently considered as candidates for black hole
mimickers~\cite{Torres:2000dw,Guzman:2005bs,AmaroSeoane:2010qx},
massive compact objects made of
axions~\cite{Hogan:1988mp,Kolb:1993zz,Barranco:2010ib}, or even the
core of the galactic halos~\cite{Sin:1992bg,Schive:2014dra} in the
context of fuzzy dark matter~\cite{Marsh:2015xka,Hui:2016ltb}.

It is within this line that we present a new class of static solutions
to the Einstein-Klein-Gordon (EKG) system which generalize the known
boson stars by considering a collection of an arbitrary odd number $N$
of free complex scalar fields of the same mass. For $N = 1$ these
configurations reduce to the standard, spherically symmetric boson
star configurations. For $N > 1$, new static configurations are
obtained which are still spherically symmetric if the collection of
scalar fields are excited in an appropriate way.

Our approach is to consider that spacetime curvature is sourced by a collection
of \emph{classical fields} that compound a spherically symmetric configuration. 
The construction is based on a method introduced in
Ref.~\cite{Olabarrieta:2007di}, where it was shown that for the case
in which the fields are real and massless, and as long as all the
harmonics with given angular momentum number $\ell = (N-1)/2$ are
excited with equal amplitude, the total stress energy-momentum tensor
is spherically symmetric. In this way, it is possible to study the
dynamics of scalar fields which, individually, have angular momentum,
while keeping the spherical symmetry of the spacetime. As remarked
in~\cite{Olabarrieta:2007di} this resembles a spherically symmetric
kinetic gas where the individual particles may rotate but the
configuration is spherical. The authors of
Ref.~\cite{Olabarrieta:2007di} used this technique to study a possible
effect of the angular momentum on the critical collapse of scalar
field configurations.

In the present work we show that their method still works for a
collection of complex, non-interacting massive scalar fields. The
situation is similar to that in standard boson stars, where an
internal $U(1)$ symmetry ``hides'' the time dependency of the field
avoiding Derrick's
theorem~\cite{Derrick:1964ww,Diez-Tejedor:2013sza}. If we extend the
group of symmetry to higher values of $N$, then not only the time
evolution, but also the angular dependency of the non-trivial
harmonics with angular momentum number $\ell=(N-1)/2$ can be
accommodated in the internal field space, leading to new static and
spherically symmetric solutions to the EKG system. We refer to these
objects as $\ell$-boson stars in this paper. The main goal of this
work is to determine their domain of existence and to compare their
properties with the standard $\ell=0$ boson stars.

The remainder of this article is organized as follows. In
Sec.~\ref{Sec:EoM} we derive the spherically symmetric field equations
for the particular collection of scalar fields belonging to a fixed
angular momentum number $\ell$. Next, in Sec.~\ref{Sec:LocSol} we
demonstrate the local existence of solutions to the EKG system in the
vicinity of the center. Based on a shooting algorithm starting from
the local solution at the center, in Sec.~\ref{Sec:GlobalSol} we
numerically solve the field equations to find the globally regular
solutions describing the $\ell$-boson stars, and discuss their
behavior, including their density profile and mass curves for
different values of $\ell$. Conclusions and an outlook to future work
are drawn in Sec.~\ref{Sec:Discussion}, and relevant identities
involving spherical harmonics that are needed in this article are
derived in the Appendix at the end of the paper.

We use the signature $(-,+,+,+)$ for the spacetime metric, and natural
units such that $\hbar=c=1$. The numerical solutions presented below
are in Planck units, where we have also taken the gravitational
constant equal to 1, $G=1$.

\section{Field equations}
\label{Sec:EoM}

In this section, we present the spherically symmetric EKG system that
describes a collection of an arbitrary odd number of complex,
non-interacting scalar fields $\Phi_i$, $i=1,\ldots, N$, of mass $\mu$
each, which are excited in an appropriate way consistent with the
spacetime symmetries.

The stress energy-momentum tensor associated with such a collection of
scalar fields is
\begin{eqnarray}\label{eq.EM}
T_{\mu\nu} = \frac{1}{2}\sum_{i}\left[\nabla_\mu\Phi_i^*\nabla_\nu\Phi_i + \nabla_\mu\Phi_i\nabla_\nu\Phi_i^*\right.\nonumber\\
 \left.- g_{\mu\nu}\left( \nabla_\alpha\Phi_i^*\nabla^\alpha\Phi_i + \mu^2\Phi_i^*\Phi_i \right)\right].
\end{eqnarray}
Here $\Phi_i^*$ denotes the complex conjugate of the field component
$\Phi_i$, and $\nabla_\mu$ is the covariant derivative with respect to
the spacetime metric $g_{\mu\nu}$. Notice that since the mass is the
same for the $N$ fields, and they do not interact between themselves,
this expression is invariant under $U(N)$ transformations in the
internal field space. We assume that the different fields in the
configuration satisfy the Klein-Gordon equation individually, such
that $(\nabla_{\mu}\nabla^{\mu}-\mu^2)\Phi_i=0$.

Following~\cite{Olabarrieta:2007di} we now consider solutions of the
form
\begin{equation}\label{eq.ansatz}
\Phi_{\ell m}(t,r,\vartheta,\varphi) = \phi_\ell(t,r) Y^{\ell m}(\vartheta,\varphi),
\end{equation}
where the angular momentum number $\ell=(N-1)/2$ is fixed in the
configuration and $m$, which plays the role of the index~$i$ in
Eq.~(\ref{eq.EM}), takes values $m = -\ell,-\ell+1,\ldots,\ell$. As
usual $Y^{\ell m}$ denotes the standard spherical harmonics, and the
amplitudes $\phi_\ell(t,r)$ are the {\it same} for all $m$.  As shown
in the Appendix this leads to a total stress energy-momentum tensor
which is spherically symmetric. (The authors of
Ref.~\cite{Olabarrieta:2007di} show a similar result valid for a
collection of scalar fields which are, however, real and massless; for
this reason we include here an independent proof.)

Representing the spacetime metric in terms of the Misner-Sharp mass
$M(t,r)$ and the lapse function $\alpha(t,r)$, the line element is
written as
\begin{equation}
ds^2 = -\alpha^2 dt^2 + \gamma^2 dr^2 + r^2 d\Omega^2,\quad
\gamma^2 := \frac{1}{1 - \frac{2M}{r}} ,
\label{Eq:Metric}
\end{equation}
with $r$ the areal radial coordinate and $d\Omega^2$ the standard line
element on the unit two-sphere.  With this notation and the assumption
in Eq.~(\ref{eq.ansatz}), the EKG system yields
\begin{subequations}\label{eq:ESS}
\begin{eqnarray}
\dot{M} &=& -\frac{\kappa_\ell r^2}{2}\frac{\alpha}{\gamma} j^{(\ell)},
\label{Eq:D0m}\\
M' &=& \frac{\kappa_\ell r^2}{2}\rho^{(\ell)},
\label{Eq:mprime}\\
\frac{\alpha'}{\alpha} &=& \gamma^2 \left( \frac{M}{r^2} + \frac{\kappa_\ell r}{2} S^{(\ell)} \right),\\
\label{Eq:D0nu}
\dot{\phi}_\ell &=& \alpha\Pi_\ell,\\
\dot{\Pi}_\ell &=& \frac{1}{r^2\gamma}\left( r^2\frac{\alpha}{\gamma} \phi_\ell' \right)' 
 + \frac{\kappa_\ell r}{2} \alpha\gamma j^{(\ell)}\Pi_\ell \nonumber\\
 &-& \alpha\left(  \mu^2 + \frac{\ell(\ell+1)}{r^2} \right) \phi_\ell,
\end{eqnarray}
\end{subequations}
where $\kappa_\ell := (2\ell+1)\: G$ is the (rescaled) gravitational
coupling constant and where the source terms are given
by\footnote{Note that when multiplied by the constant factor
  $(2\ell+1)/(8\pi)$, these quantities correspond to the energy
  density, $\rho=n^{\mu}n^{\nu}T_{\mu\nu}$, the radial stress
  $S=P^{\mu}_{r}P^{\nu}_{r}T_{\mu\nu}$, and the radial momentum flux
  $j=-P^{\mu}_{r}n^{\nu}T_{\mu\nu}$, respectively, of matter as
  measured in a local orthonormal frame by the Eulerian observers
  moving along the direction normal to the spatial hypersurfaces,
  where $n^{\mu}$ is the unit normal vector to these hypersurfaces,
  and $P^{\mu}_{r}$ the orthogonal projection operator onto them.}
\begin{subequations}\label{eq:source}
\begin{eqnarray}
\rho^{(\ell)} &=& |\Pi_\ell|^2 + |\chi_\ell|^2 + \left( \mu^2 + \frac{\ell(\ell+1)}{r^2} \right) |\phi_\ell|^2 ,
\label{eq:den_flux}\\
S^{(\ell)} &=& |\Pi_\ell|^2 + |\chi_\ell|^2 - \left( \mu^2 + \frac{\ell(\ell+1)}{r^2} \right) |\phi_\ell|^2,\\
j^{(\ell)} &=& -2\re( \Pi_\ell^*\chi_\ell ). 
\end{eqnarray}
\end{subequations}
Here $\Pi_\ell := \alpha^{-1}\dot{\phi}_\ell$ is the conjugate
momentum associated with the scalar field, $\chi_\ell :=
\gamma^{-1}\phi_\ell'$, and a dot and a prime denote partial
derivatives with respect to $t$ and $r$, respectively.

In the following, we look for solutions with a harmonic
time-dependency,
\begin{equation}\label{eq.harmonic}
\phi_\ell(t,r) = e^{i\omega t}\psi_\ell(r),
\end{equation}
with some real frequency $\omega$ (taken to be positive without loss
of generality) and a real-valued function $\psi_\ell(r)$. This
provides a nonlinear eigenvalue problem for the quantities
$(M,\alpha\gamma,\psi_\ell)$ which can be written as
\begin{subequations}\label{Eq:bosonstars}
\begin{eqnarray}
&& M' = \frac{\kappa_\ell r^2}{2}
\left[ \frac{\psi_\ell'^2}{\gamma^2}
 + \left(\mu^2 + \frac{\omega^2}{\alpha^2} + \frac{\ell(\ell+1)}{r^2} \right)\psi_\ell^2 \right],\quad
\label{Eq:bosonstars.2} \\
&& \frac{(\alpha\gamma)'}{\alpha\gamma^3} 
 = \kappa_\ell r\left[ \frac{\psi_\ell'^2}{\gamma^2} +  \frac{\omega^2}{\alpha^2}\psi_\ell^2 \right],
\label{Eq:bosonstars.3} \\
&& \frac{1}{r^2\alpha\gamma}\left( \frac{r^2\alpha}{\gamma}\psi_\ell' \right)' 
 = \left(\mu^2 - \frac{\omega^2}{\alpha^2} + \frac{\ell(\ell+1)}{r^2} \right)\psi_\ell.
\label{Eq:bosonstars.1}
\end{eqnarray}
\end{subequations}
For the particular case of $N=1$, i.e. $\ell = 0$, these equations
reduce to the ones describing static boson stars with a zero angular
momentum scalar field. These configurations have been studied
extensively in the literature, and may be parametrized by the central
value of the field and an excitation
number~\cite{Jetzer:1991jr,Lee:1991ax,Schunck:2003kk,Liebling:2012fv}.

In the following, we study the solutions for $N>1$, i.e. $\ell >
0$. In this case the presence of the centrifugal term
$\ell(\ell+1)/r^2$ in the EKG system requires $\psi_\ell$ to vanish at
the center of the configuration and to decay with a specific rate as
$r\to 0$. This is analyzed next.

\section{Local solutions near the center}
\label{Sec:LocSol}

For $\ell = 0$, a full proof for the existence of regular solutions of
finite mass was given by Bizo\'n and Wasserman in
Ref.~\cite{Bizon:2000es}. Although it would be very interesting to
generalize their results to arbitrary values of $\ell$, a full
existence proof clearly lies beyond the scope of this work. However,
for the purpose of this paper we found it useful to generalize one
partial result given in~\cite{Bizon:2000es} (see Proposition~3.1 of
that paper), regarding the local existence of solutions near the
center. This provides a rigorous discussion for the existence and
behavior of the solutions in the vicinity of $r = 0$, and allows us to
set up the correct boundary conditions to start the numerical
integration scheme described in the next section.

In order to discuss the solutions in the vicinity of $r = 0$, it is
convenient to introduce the following dimensionless
quantities:\footnote{In~\cite{Bizon:2000es} the quantities $A :=
  \gamma^{-2}$ and $C := \gamma^2/Q$ are used instead of $\gamma$ and
  $Q$.}
\begin{subequations}
\begin{eqnarray}
 x := \mu r,&&\;
f(x) := \sqrt{\kappa_\ell}\psi_\ell(r),\\
 B(x) := 2\mu M(r)/x^2,&&\;
Q(x) := \frac{\mu}{\omega}\alpha(r)\gamma(r),
\end{eqnarray}
\end{subequations}
in terms of which the system~(\ref{Eq:bosonstars}) can be rewritten as
\begin{subequations}\label{Eq:bosonstarsBis}
\begin{eqnarray}
&& \frac{d}{dx}(x^2 B) = x^2
\left[ \frac{f_x^2}{\gamma^2}
 + \left( 1 + \frac{\gamma^2}{Q^2} + \frac{\ell(\ell+1)}{x^2} \right) f^2 \right],\qquad
\label{Eq:bosonstars.2Bis} \\
&& \frac{d}{dx}\log Q = x\left[ f_x^2 +  \frac{\gamma^4}{Q^2} f^2 \right],
\label{Eq:bosonstars.3Bis} \\
&& \frac{1}{x^2 Q}\frac{d}{dx}\left( \frac{x^2 Q}{\gamma^2} f_x \right) 
 = \left( 1 - \frac{\gamma^2}{Q^2} + \frac{\ell(\ell+1)}{x^2} \right) f.
\label{Eq:bosonstars.1Bis}
\end{eqnarray}
\end{subequations}
Here we have used the shorthand notation $f_x = df/dx$, and it is
understood that $\gamma^2$ should be substituted by $[ 1 - x
  B(x)]^{-1}$.

For $\ell = 0$, it was shown in Ref.~\cite{Bizon:2000es} that for each
positive values of $Q_0$ and $a_0$, there exists a local solution of
the form
\begin{subequations}\label{Eq:AsymptoticL=0}
\begin{eqnarray}
B(x) &=& {\cal O}(x),\\
Q(x) &=& Q_0 + {\cal O}(x^2),\\
f(x) &=& a_0 + {\cal O}(x^2),
\end{eqnarray}
\end{subequations}
near the center $x = 0$. In the following, we generalize this local
existence result to arbitrary values of $\ell$. As mentioned above,
the presence of the centrifugal term $\ell(\ell+1)/x^2$ drastically
modifies the behavior of the fields close to the center. In fact, a
heuristic approximation assuming $B\to 0$, $Q\to \textrm{const.}$
gives
\begin{equation}
-\frac{d}{dx}(x^2 f_x) + \ell(\ell+1) f \simeq 0,
\end{equation}
which has solutions of the form $f\sim x^\ell$ or $f\sim
x^{-\ell-1}$. Since we require regularity a the center, we discard the
second possibility. We now show the existence of a two-parameter
family of solutions for which
\begin{subequations}\label{Eq:AsymptoticL>0}
\begin{eqnarray}
B(x) &=& {\cal O}(x^{2\ell-1}),\\
Q(x) &=& Q_0\left[ 1 + {\cal O}(x^{2\ell}) \right],\\
f(x) &=& \frac{a_0}{2\ell+1} x^{\ell}\left[ 1 + {\cal O}(x^2) \right].
\end{eqnarray}
\end{subequations}
This implies $\alpha = \alpha_c[ 1 + {\cal O}(x^{2\ell}) ]$, with
$\alpha_c = Q_0\omega/\mu$, and $M = {\cal O}(x^{2\ell+1})$, in terms
of the original variables.  Notice that these expressions are only
valid as long as $\ell>0$.

In order to prove the existence of such solutions, we replace $f$ and
its first derivative with the new fields $(G,H)$, such that
\begin{equation}
\left( \begin{array}{r} f(x) \\ x f_x(x) \end{array} \right)
 = \frac{x^\ell}{2\ell+1} \left( \begin{array}{cc} 1 & 1 \\ \ell & -(\ell+1) \end{array} \right)
 \left( \begin{array}{r} G(x) \\ H(x) \end{array} \right).
\label{Eq:GHDef}
\end{equation}
Note that (up to a constant $2\ell+1$) the two asymptotic solutions
$f(x) = x^\ell$ and $f(x) = x^{-\ell-1}$ correspond to the fields
$(G,H) = (1,0)$ and $(G,H) = x^{-2\ell-1}(0,1)$, respectively. In
terms of the fields $u := (B,Q,G,H)$, the
system~(\ref{Eq:bosonstarsBis}) can be rewritten in first-order form:
\begin{subequations}\label{Eq:FOM}
\begin{eqnarray}
\frac{d}{dx} (x^2 B) &=& x^{2\ell} F_B(x,u),\\
\frac{dQ}{dx} &=& x^{2\ell-1} F_Q(x,u),\\
\frac{dG}{dx} &=& F_G(x,u),\\
\frac{d}{dx}(x^{2\ell+1} H) &=& -x^{2\ell+1} F_G(x,u),
\end{eqnarray}
\end{subequations}
with the source terms given by
\begin{eqnarray}
F_B &=& \frac{(x^{1-\ell} f_x)^2}{\gamma^2}
 + \left[ x^2\left(1 + \frac{\gamma^2}{Q^2}\right) + \ell(\ell+1) \right]
 (x^{-\ell} f)^2, \nonumber \\
F_Q &=& Q(x^{1-\ell} f_x)^2  + \frac{\gamma^4}{Q} x^2 (x^{-\ell} f)^2, \\
F_G &=& \gamma^2 \left\{ \left[ x\left( 1 - \frac{\gamma^2}{Q^2} \right) + \ell(\ell+1) B \right]
 (x^{-\ell} f) \right.  \nonumber \\
&&  + \left[ x^{2\ell-1}\left( x^2 + \ell(\ell+1) \right) (x^{-\ell} f)^2 - B
\right] (x^{1-\ell} f_x) \Big\}. \nonumber
\end{eqnarray}
In these expressions, $\gamma^2$ should be replaced with $[1 -
  xB]^{-1}$ and the terms $x^{-\ell} f$ and $x^{1-\ell} f_x$ should be
rewritten in terms of $G$ and $H$ according to
Eq.~(\ref{Eq:GHDef}). Note that when rewritten in this way, these
terms are regular at $x = 0$, which implies that the source terms
$F_B$, $F_Q$ and $F_G$ are regular as well at $x = 0$ for all $\ell
\geq 1$.

Eqs.~(\ref{Eq:FOM}) can be reformulated as a fixed point problem of
the form $u = T u$, with the nonlinear map $T u = (T_1 u,T_2 u,T_3
u,T_4 u)$ given by
\begin{subequations}
\begin{eqnarray}
T_1 u(x) &:=& x^{-2}\int_0^x y^{2\ell} F_B(y,u(y)) dy,\\
T_2 u(x) &:=& Q_0 + \int_0^x y^{2\ell-1} F_Q(y,u(y)) dy,\\
T_3 u(x) &:=& a_0 + \int_0^x F_G(y,u(y)) dy,\\
T_4 u(x) &:=& x^{-2\ell-1}\int_0^x y^{2\ell+1} F_G(y,u(y)) dy.
\end{eqnarray}
\end{subequations}
Introducing the function space $X$ of continuous, bounded functions
$u$ on some interval $[0,x_0]$ with the $\sup$-norm, and the closed
subspace $Y\subset X$ consisting of those fields $u\in X$ whose
distance to the point $u_0 := ( 0,Q_0,a_0,0)$ is no greater than $1$,
it is not difficult to show that $T$ leaves $Y$ invariant and defines
a contraction on $Y$, provided $x_0$ is small enough. According to the
contraction mapping principle, $T$ possesses a unique fixed point in
$Y$, and this fixed point provides the required solution. The solution
may be constructed by means of the iteration $u_0$, $u_1 := T u_0$,
$u_2 := T u_1$, $\ldots$, from which one easily finds the asymptotic
behavior~(\ref{Eq:AsymptoticL>0}).

\section{Global solutions}
\label{Sec:GlobalSol}

\begin{figure*}
\begin{overpic}[width=0.48\textwidth]{MvsR99_all.eps}  
\put(75.6,65){\includegraphics[scale=0.008]{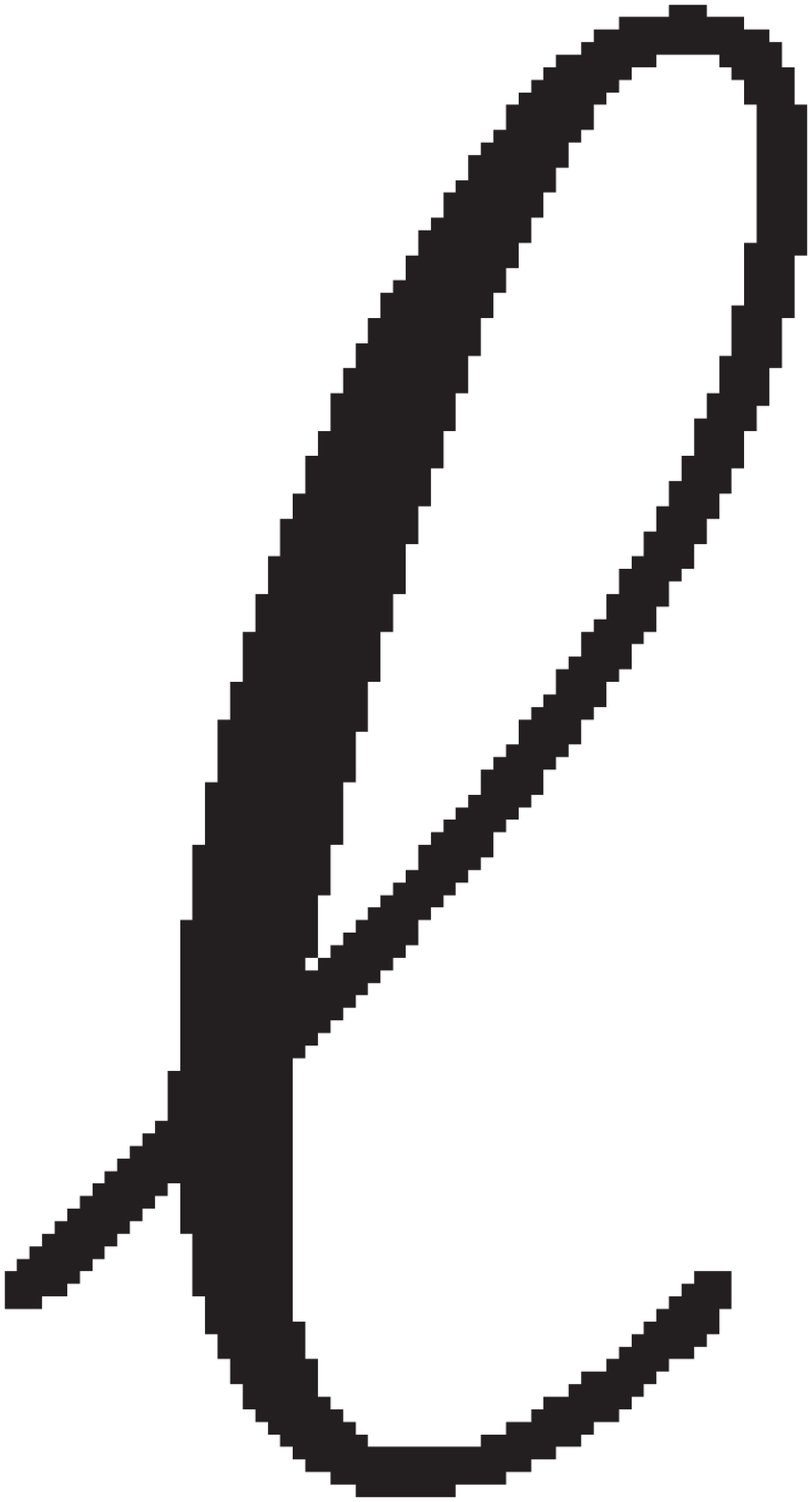}}
\put(75.6,61.2){\includegraphics[scale=0.008]{ell.eps}}
\put(75.6,57.9){\includegraphics[scale=0.008]{ell.eps}}
\put(75.6,54.7){\includegraphics[scale=0.008]{ell.eps}}
\put(75.6,51.5){\includegraphics[scale=0.008]{ell.eps}}
\end{overpic}\qquad
\begin{overpic}[width=0.48\textwidth]{MvsOmega.eps}
\put(75.4,64.6){\includegraphics[scale=0.008]{ell.eps}}
\put(75.4,60.8){\includegraphics[scale=0.008]{ell.eps}}
\put(75.4,57.6){\includegraphics[scale=0.008]{ell.eps}}
\put(75.4,54.4){\includegraphics[scale=0.008]{ell.eps}}
\put(75.4,51.2){\includegraphics[scale=0.008]{ell.eps}}
\end{overpic}
\caption{{\it Left panel:} Total mass vs effective radius for
  equilibrium configurations of different angular momentum number
  $\ell$.  {\it Right panel:} Total mass vs frequency of oscillation
  for the same configurations as in the left panel.}
 \label{fig:1}
\end{figure*}

In this section, we present numerical solutions to the EKG system
described by Eqs.~(\ref{Eq:bosonstars}), that extend the local
solutions of the previous section to provide global configurations
that are regular everywhere and posses a finite mass. From this point
onwards, and in order to simplify the numerical analysis, we set
$G=1$, such that all quantities are dimensionless and measured in
Planck units.

To proceed, we solve numerically the EKG system~(\ref{Eq:bosonstars})
expressed in the alternative form
\begin{widetext}
\begin{subequations}\label{eqs.numerical}
\begin{eqnarray}
\gamma'=\frac{2\ell+1}{2}r\gamma\left[ \left( 
\frac{\omega^2}{\alpha^2}+\frac{\ell(\ell+1)}{r^2}
+\mu^2\right)\gamma^2u_{\ell}^2r^{2\ell}+(u_{\ell}'r^{\ell}+\ell u_{\ell} r^{\ell-1})^2\right]-\left(\frac{
\gamma^2-1}{2r}\right)\gamma,\\
\alpha'=\frac{2\ell+1}{2}r\alpha\left[ \left( 
\frac{\omega^2}{\alpha^2}-\frac{\ell(\ell+1)}{r^2}
-\mu^2\right)\gamma^2u_{\ell}^2r^{2\ell}+(u_{\ell}'r^{\ell}+\ell u_{\ell} r^{\ell-1})^2\right]+\left(\frac{
\gamma^2-1}{2r}\right)\alpha,\\
u_{\ell}''=\left( \mu^2-\frac{\omega^2}{\alpha^2}\right)\gamma^2 u_{\ell}
-\left( \gamma^2+2\ell+1 \right)\frac{u_\ell'}{r}
+\ell^2\left( \gamma^2-1\right)\frac{u_{\ell}}{r^2}
+(2\ell+1)\left (\mu^2+\frac{\ell(\ell+1)}{r^2}\right)\gamma^2\left(ru'_{\ell}+\ell u_{\ell} \right)u_{\ell}^2r^{2\ell} ,
\end{eqnarray}
\end{subequations}
\end{widetext}
where $u_{\ell}:=\psi_{\ell}/r^{\ell}$.  The choice of appropriate
boundary conditions must guarantee that the solutions are regular and
asymptotically flat. Demanding regularity at the origin, i.e.
\begin{subequations}\label{eq.conditions1}
\begin{eqnarray}
u_{\ell}(r=0)&=&u_{\ell}^0,\\
u_{\ell}'(r=0)&=&0,\\
\alpha(r=0)&=&1,\\
\gamma(r=0)&=&1
\end{eqnarray}
\end{subequations}
(see Sec.~\ref{Sec:LocSol} for details), and a vanishing field
amplitude at infinity, i.e. $u_\ell(r\to\infty)= 0$, one obtains a
nonlinear eigenvalue problem for the mode-frequency $\omega$. Here
$u_{\ell}^0$ is an arbitrary positive constant, and with no loss of
generality we have fixed the value of the lapse function at the origin
to one. Notice that since the system of equations is invariant under
the transformation $(\alpha, \omega)\mapsto\lambda (\alpha, \omega)$,
with some positive arbitrary constant $\lambda$, one can always
rescale the value of the lapse function at the end of the numerical
integration in such a way that $\alpha(r\to\infty)=1$, providing an
asymptotic flat coordinate system at infinity. We always perform such
a rescaling when reporting our results below.

A word about our numerical algorithm is on order here. The integration
of the system is performed using a shooting algorithm to find the
frequencies $\omega$. Notice that, as mentioned above, we have
rescaled the field as $u_{\ell}:=\psi_{\ell}/r^{\ell}$, and we solve
for $u_{\ell}$ instead of $\psi_{\ell}$. The reason for this is that,
for $\ell>1$, one finds that $\psi_{\ell}$ and its first $\ell-1$
derivatives vanish at the origin, which results in a numerically ill
behaved set of boundary conditions. On the other hand, $u_{\ell}$ has
a constant value at the origin and only its first derivative vanishes,
which results in a numerically well behaved system of equations plus
boundary conditions.

To find the solution one then integrates the system of
Eqs.~(\ref{eqs.numerical}) outwards from the origin, with initial
conditions given by Eqs.~(\ref{eq.conditions1}), and searches for the
values of the frequency that match the required asymptotic behavior of
the field (the solution should decay exponentially), until the
shooting parameter converges to the desired accuracy. Notice that this
process is somewhat delicate as for arbitrary $\omega$ there is always
a solution that grows exponentially, and the shooting parameter must
be precise enough to eliminate the growing solution.  In practice we
start with the outer boundary fairly close in, find a good initial
guess for $\omega$, and then slowly move the boundary outward,
adjusting the value of $\omega$ to high numerical precision as we do
so.  We have in fact constructed two independent codes, with a fourth
order and a fifth order Runge-Kutta numerical integration, and we have
checked that the solutions of both codes converge with numerical
resolution as expected, and that these solutions match.

For simplicity, in all our solutions we take the mass parameter
\mbox{$\mu=1$}, although the solutions can be rescaled to an arbitrary
value of $\mu$ using the invariance of the equation system under the
transformation
\begin{equation}
 \mu \mapsto \lambda \mu,\quad \omega\mapsto \lambda \omega,
\quad r\mapsto \lambda^{-1} r,
\quad u_{\ell} \mapsto \lambda^{\ell}u_{\ell},
\end{equation}
with the metric coefficients $\alpha$ and $\gamma$ unchanged.

We characterize the total mass of an $\ell$-boson star in terms of the
asymptotic value of the Misner-Sharp mass function, which is
approximated by evaluating the metric coefficient $\gamma(r)$ at the
last grid point $r_{\rm{max}}$ of the computational domain, such that
\begin{equation}
M \approx \frac{r_{\textrm{max}}}{2}\left[1-\frac{1}{\gamma^2(r_{\rm{max}})}\right].
\end{equation}
Although $\ell$-boson stars extend to infinity and do not posses a
surface at a finite radius like usual fluid stars, one can associate
to them an effective radius $R(99\%)$ defined as the areal radius of
the object which contains $99\%$ of the total mass.

For a given angular momentum number $\ell$, the equilibrium
configurations are labeled by a continuous parameter representing the
field amplitude, i.e. the constant $u_{\ell}^0$ in
Eq.~(\ref{eq.conditions1}), and a discrete number $n$ that labels the
solutions to the eigenvalue problem. In this paper we restrict our
attention to ground state configurations only, $n=1$, corresponding to the
solutions with the lowest possible value of the frequency $\omega$ for
a given angular momentum number and field amplitude.

Each of these configurations have an associated frequency $\omega$, a
total mass $M$, and an effective radius $R(99\%)$. In Fig.~\ref{fig:1}
we show, for the equilibrium configurations with $\ell=0,1,2,3,4$, a
plot for the total mass as a function of the effective radius (left
panel) and frequency (right panel), respectively. Notice that the mass
increases monotonically with the effective radius up to a maximum
value, and then decreases, in analogy with what is known for standard
$\ell=0$ boson stars. The configurations that correspond to this
maximum value of the mass for the different values of $\ell$ have been
labeled as A, B, C, D, and E in the figure. The main characteristics
of those configurations are shown in Table \ref{Table:1}. Notice also
that the compactness of these objects, defined as the ratio of the
total mass to the effective radius, grows with the angular momentum
number, at least for the first five possible values of $\ell$
considered in this paper. The relation of the mass with the frequency
resembles some of the characteristics of rotating boson stars. Notice
that since the centrifugal term decreases with increasing areal
radius, the global characteristics of these objects approach those of
standard bosons stars when they grow in size, the differences being
only close to the center of the configuration.

\begin{figure*}
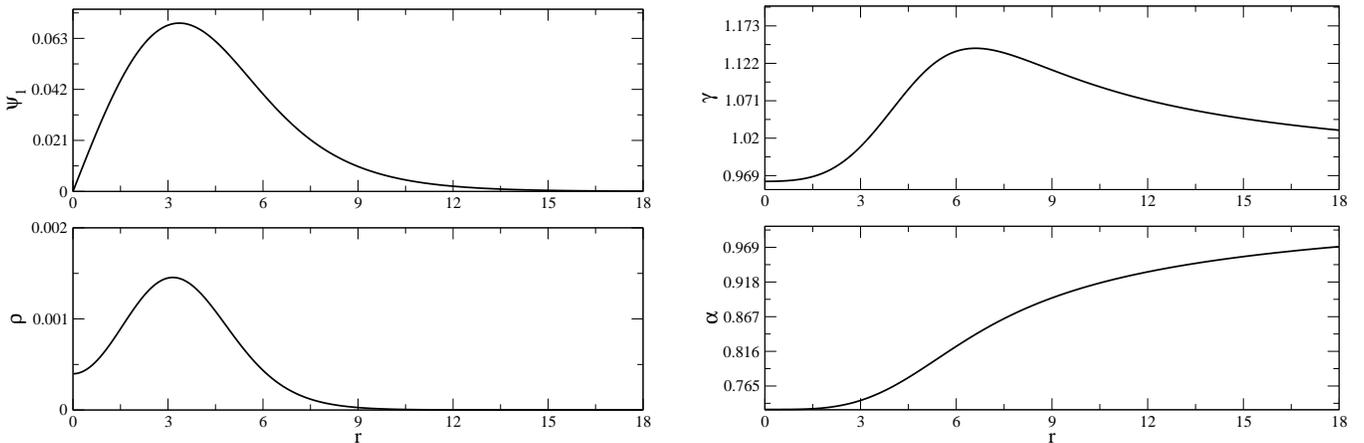

 \includegraphics[width=0.48\textwidth]{psi_l1_max.eps}\qquad  
 \includegraphics[width=0.48\textwidth]{metrica_l1_max.eps}
 \caption{Self-gravitating $\ell$-boson star corresponding to point B
   in Fig.~\ref{fig:1}. {\it Left panel:} wave-function $\psi_1(r)$,
   and energy density $\rho(r)=(2\ell+1)\rho^{(\ell)}(r)/(8\pi)$, as
   functions of the radial coordinate. {\it Right panel:} the metric
   coefficients $\gamma(r)$ and $\alpha(r)$, also as a function of the
   radial coordinate. Notice that even though the wave-function
   vanishes at the origin, the energy density takes a value different
   from zero at that point. This is a characteristic of the $\ell=1$
   configurations. See Fig.~\ref{fig.energydensity} for energy density
   profiles in the cases of $\ell=0$, 1 and 2. The regularity of these
   objects at the origin in evident in the profiles for the metric
   coefficients.}
 \label{fig:pointB}
\end{figure*}

\begin{figure*}
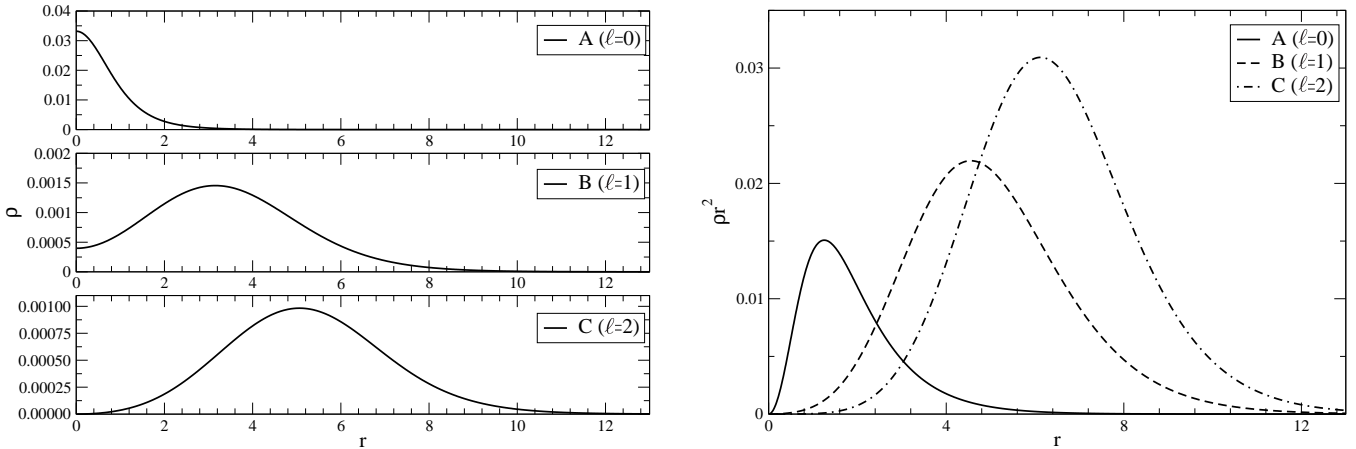

\begin{overpic}[width=0.48\textwidth]{rho.eps}
\put(92.7,62.2){\includegraphics[scale=0.008]{ell.eps}}
\put(92.7,39.8){\includegraphics[scale=0.008]{ell.eps}}
\put(92.7,17.5){\includegraphics[scale=0.008]{ell.eps}}
\end{overpic}\qquad
\begin{overpic}[width=0.48\textwidth]{r2density.eps}
\put(92.3,62.2){\includegraphics[scale=0.008]{ell.eps}}
\put(92.3,58.5){\includegraphics[scale=0.008]{ell.eps}}
\put(92.3,54.9){\includegraphics[scale=0.008]{ell.eps}}
\end{overpic}
\caption{ {\it Left panel:} Profile of the energy density
  $\rho(r)=(2\ell+1)\rho^{(\ell)}(r)/(8\pi)$ for the configurations A,
  B, and C in Fig.~\ref{fig:1}, respectively. Notice that the vertical
  scale changes between the plots. It is interesting to emphasize that
  $\rho(0)\neq 0$ if $\ell=0$ or 1, but $\rho(0)= 0$ if $\ell>1$. Only
  if $\ell=0$ the energy density takes its maximum value at the
  origin.  {\it Right panel:} Same profiles as in left panel, but
  multiplied by the square of the areal radius. Even if the amplitude
  of the energy density decreases as the angular momentum number
  increases, the configurations become more extended, and then the
  combination $\rho r^2$ grows, leading to more massive objects; see
  Table~\ref{Table:1} for details.}
 \label{fig.energydensity}
\end{figure*}

A representative $\ell$-boson star configuration appears in
Fig.~\ref{fig:pointB}, where we show, for the configuration~B of
Fig.~\ref{fig:1}, the profile of the wave-function $\psi_1(r)$,
introduced in Eq.~(\ref{eq.harmonic}), the energy density
$\rho(r)=(2\ell+1)\rho^{(\ell)}(r)/(8\pi)$, where $\rho^{(\ell)}$ was
given in Eq.~(\ref{eq:den_flux}), and the metric coefficients
$\gamma(r)$ and $\alpha(r)$ of Eq.~(\ref{Eq:Metric}), as functions of
the areal radius. Notice that, contrary to what happens for the
standard $\ell=0$ boson stars and perfect fluid configurations, the
energy density at the origin does not take a maximum value that
decreases monotonically to zero when the angular momentum number is
non-trivial. This is a consequence of the regularity conditions
discussed in Sec.~\ref{Sec:LocSol}. In particular, the energy density
takes a local minimum at the origin for the configurations with
$\ell=1$, and vanishes if $\ell>1$. See Fig.~\ref{fig.energydensity}
for details.

\begin{table}
\begin{tabular}{c|c|c|c|c}
\hline
Configuration & $M$ &  $\;R(99\%)\;$ & $\omega$ & $M/R(99\%)$\\
\hline
A ($\ell=0$)  & $\;0.63\;$  & $7.89$  & $\;0.854\;$ &  $0.08$\\
B ($\ell=1$)  & $1.18$  & $12.75$ & $0.836$ &  $0.09$\\
C ($\ell=2$)  & $1.72$  & $15.35$ & $0.832$ &  $0.11$\\
D ($\ell=3$)  & $2.25$  & $17.22$ & $0.820$ &  $0.13$\\
E ($\ell=4$)  & $2.78$  & $19.80$ & $0.819$ &  $0.14$\\
\hline
\end{tabular}
\caption{Main characteristics of some $\ell$-boson stars with
  different angular momentum number, reported in Planck units. The
  labels A, B, C, D and E make reference to those in
  Fig.~\ref{fig:1}. Notice that the objects with a non-trivial angular
  momentum number are more compact than standard boson stars with
  $\ell=0$.}
\label{Table:1}
\end{table}

\section{Discussion and conclusions}
\label{Sec:Discussion}

We have introduced new solutions to the static EKG system of equations
that describe regular self-gravitating objects of finite mass. These
configurations incorporate some of the effects of angular momentum
while keeping the spherical symmetry of the spacetime metric,
generalizing standard non-rotating boson stars.  The technique
requires an arbitrary odd number $N$ of massive, complex,
non-interacting scalar fields with an internal $U(N)$ symmetry, and
are labeled by an angular momentum number, $\ell=(N-1)/2$, and the
usual field amplitude and excitation number. This motivates the name
$\ell$-boson star. In this paper we have restricted our attention to
ground state configurations only, although this can be easily
generalized to the case of excited states.

We have shown that these objects possess similar properties to those
of standard $N=1$, $\ell=0$, boson stars. Specifically, for a given
angular momentum number, the equilibrium configurations exhibit a
maximum value of the mass, and this maximum grows as $\ell$ increases,
leading to more compact objects. For the case of $\ell=0$, it is well
known that the maximum mass configuration divides the stable and
unstable branches. We have done some preliminary numerical evolutions
for the case with $\ell=1$ and have found that the same appears to be
true: for a given mass below the maximum, configurations with a large
effective radius and higher frequencies (those to the right of the
maximum in both panels of Fig.~\ref{fig:1}) remain stable when
perturbed, while those with smaller effective radius and smaller
frequencies (to the left of the maximum in Fig.~\ref{fig:1}) either
collapse rapidly to a black hole, or oscillate with a very long period
around a less compact state (depending on the precise way in which
they are perturbed).  We will report on the results of these
simulations and those for $\ell>1$ elsewhere.

Even if in this paper we restricted our attention to the ground state $n=1$ and
only one single value of $\ell$ at a time in the configurations,
the system of Eqs.~(\ref{Eq:bosonstars}) can easily be
generalized to the case when several different modes $(n,\ell)$ are present
simultaneously. This can be achieved by summing over all the excited states in
the right hand side of Eqs.~(\ref{Eq:bosonstars.2}) and~(\ref{Eq:bosonstars.3}), similarly
to the multi-state configurations constructed in~\cite{Bernal:2009zy,UrenaLopez:2010ur} for $\ell=m=0$. 
Under the classical approach that we assume in this paper, this
requires the inclusion of more fields, making it possible to use
not only odd, but also even values of $N$, and to obtain more general
density profiles than those reported here. We leave a more detailed
analysis for a companion paper.

The main implication of this work is that the solution space of
spherically symmetric, static boson stars is much larger and has a
much richer structure when a collection of scalar fields is considered
instead of a single complex scalar field that necessarily requires of
the harmonic \mbox{$\ell=m=0$}. An alternative interpretation for the
collection of scalar fields considered here, 
and potential applications for dark matter observations, will be explored in another
paper.


\acknowledgments
This work was supported in part by CONACYT grants No. 82787, 101353, 259228,
167335, and 182445, by the CONACyT Network Project No. 294625 ``Agujeros Negros y Ondas Gravitatorias", by CONACYT Fronteras Project 281, by DGAPA-UNAM through 
grants IN110218 and IA101318, by SEP-23-005 through grant 181345, and by a CIC grant to Universidad Michoacana de San Nicol\'as de Hidalgo. A.B. and A.D.-T. were 
partially supported by PRODEP.

\appendix

\section{Stress energy-momentum tensor for the collection of scalar fields given in Eq.~(\ref{eq.ansatz})}
\label{App:StreeEnergyT}

The purpose of this appendix is to provide a proof for the statement regarding the fact that the stress energy-momentum tensor associated with a collection
of scalar fields of the form in Eq.~(\ref{eq.ansatz}) is spherically symmetric. For an alternative proof based on methods from quantum mechanics, see the Appendix in 
Ref.~\cite{Olabarrieta:2007di}. 

The proof here is based directly on the well-known identity
\begin{equation}
\sum\limits_{m=-\ell}^\ell |Y^{\ell m}(\vartheta,\varphi)|^2 = \frac{2\ell+1}{4\pi},
\label{Eq:AdditionTheorem}
\end{equation}
which upon differentiation and taking into account that $(Y^{\ell m})^* = Y^{\ell -m}$, yields
\begin{equation}
\sum\limits_{m=-\ell}^\ell Y^{\ell m}(\vartheta,\varphi)^* \hat{\nabla}_A Y^{\ell m}(\vartheta,\varphi) = 0.
\label{Eq:AdditionTheoremDeriv1}
\end{equation}
Here and in the following, $\hat{\nabla}_A$ and $\hat{g}_{AB}$ refer to the covariant derivative and standard metric, respectively, on the unit two-sphere. 
Taking a further derivative gives
\begin{equation}
\sum\limits_{m=-\ell}^\ell ( \hat{\nabla}_A Y^{\ell m} )^*( \hat{\nabla}_B Y^{\ell m} ) 
 = -\sum\limits_{m=-\ell}^\ell (Y^{\ell m})^* \hat{\nabla}_A\hat{\nabla}_B Y^{\ell m}.
\label{Eq:AdditionTheoremDeriv2}
\end{equation}
Since $\hat{\nabla}^A\hat{\nabla}_A Y^{\ell m} = -\ell(\ell+1) Y^{\ell m}$, the trace of this equation together with Eq.~(\ref{Eq:AdditionTheorem}) provide the 
following identity:
\begin{equation}
\sum\limits_{m=-\ell}^\ell ( \hat{\nabla}^A Y^{\ell m} )^*( \hat{\nabla}_A Y^{\ell m} ) 
 = \frac{\ell(\ell+1)(2\ell+1)}{4\pi}.
\label{Eq:AdditionTheoremDeriv3}
\end{equation}

On the other hand, taking the divergence on both sides of Eq.~(\ref{Eq:AdditionTheoremDeriv2}), and using Eq.~(\ref{Eq:AdditionTheoremDeriv1}), we obtain
\begin{equation}
\sum\limits_{m=-\ell}^\ell (\hat{\nabla}^A Y^{\ell m})^* \hat{\nabla}_A\hat{\nabla}_B Y^{\ell m} = 0.
\end{equation}
The previous formula implies that the symmetric trace-free tensor field on the two-sphere
\begin{eqnarray}
\tau_{AB}
 &:=& \sum\limits_{m=-\ell}^\ell \left[ ( \hat{\nabla}_A Y^{\ell m} )^*( \hat{\nabla}_B Y^{\ell m} ) \right.\nonumber\\
 && \left. \qquad - \frac{1}{2}\hat{g}_{AB}(\hat{\nabla}^C Y^{\ell m})^*(\hat{\nabla}_C Y^{\ell m}) \right]
\end{eqnarray}
is divergence-free. However, all symmetric, trace- and divergence-free tensor fields on the two-sphere are trivial (see, for instance Lemma~1 in the Appendix of 
Ref.~\cite{eCnOoS13}), such that $\tau_{AB} = 0$. This implies
\begin{equation}
\sum\limits_{m=-\ell}^\ell ( \hat{\nabla}_A Y^{\ell m} )^*( \hat{\nabla}_B Y^{\ell m} ) 
 = \frac{\ell(\ell+1)(2\ell+1)}{8\pi} \hat{g}_{AB}.
\label{Eq:AdditionTheoremDeriv4}
\end{equation}

Based on the identities~(\ref{Eq:AdditionTheorem}--\ref{Eq:AdditionTheoremDeriv4}), one easily finds the following expressions for the components of the stress 
energy-momentum tensor given in Eq.~(\ref{eq.EM}):
\begin{eqnarray}
T_{ab} &=& \frac{2\ell+1}{4\pi}\left[ (\partial_a\phi_\ell)^*(\partial_b\phi_\ell) 
 - \frac{1}{2}\tilde{g}_{ab}\left( \tilde{g}^{cd}(\partial_c\phi_\ell)^*(\partial_d\phi_\ell) \right.\right.\nonumber\\
 && \qquad\qquad \left. \left. + \frac{\ell(\ell+1)}{r^2} |\phi_\ell|^2 + \mu^2 |\phi_\ell|^2 \right) \right],\nonumber\\
T_{aB} &=& 0,\\
T_{AB} &=& -\frac{2\ell+1}{8\pi} r^2
\left[ \tilde{g}^{cd}(\partial_c\phi_\ell)^*(\partial_d\phi_\ell) + \mu^2 |\phi_\ell |^2 \right]\hat{g}_{AB},\nonumber
\end{eqnarray}
where here $a,b = t,r$, $A,B = \vartheta,\varphi$ and $\tilde{g}_{ab}$ refers to the components of the time--radial part $-\alpha^2 dt^2 + \gamma^2 dr^2$ of the metric.



\begin{thebibliography}{99}


\bibitem{Kaup68} D.J.~Kaup, ``Klein-Gordon geon,'' 
Phys. Rev. {\bf 172}, 1331 (1968)

\bibitem{Ruffini:1969qy}
R.~Ruffini and S.~Bonazzola, ``Systems of selfgravitating particles in general relativity and the concept of an equation of state,'' 
Phys. Rev. {\bf 187}, 1767-1783 (1969)

\bibitem{Colpi:1986ye} M.~Colpi, S.L.~Shapiro and I.~Wasserman, ``Boson stars: Gravitational equilibria of selfinteracting scalar fields,''
Phys. Rev. Lett. {\bf 57}, 2485-2488 (1986)

\bibitem{Friedberg87} R.~Friedberg, T.D.~Lee and Y.~Pang, ``Mini-soliton stars,'' 
Phys. Rev. D {\bf 35}, 3640-3657 (1987)

\bibitem{Gleiser:1988rq} M.~Gleiser, ``Stability of boson stars,''
Phys. Rev. D {\bf 38}, 2376 (1988)

\bibitem{Lee:1988av}
T.D.~Lee and Y.~Pang, ``Stability of mini-boson stars,''
Nucl. Phys. B {\bf 315}, 477 (1989)

\bibitem{Seidel:1990}
E.~Seidel and W-M.~Suen, "Dynamical evolution of boson stars: Perturbing the ground state,"
Phys. Rev. D {\bf 42}, 384 (1990)

\bibitem{Guzman:2004wj} F.S.~Guzman and L.A.~Urena-Lopez, ``Evolution of the Schrodinger-Newton system for a selfgravitating scalar field,''
Phys. Rev. D {\bf 69}, 124033 (2004) [arXiv:0404014 [gr-qc]]

\bibitem{Jetzer:1991jr} P.~Jetzer, ``Boson stars,''
Phys. Rept. {\bf 220}, 163-227 (1992)

\bibitem{Lee:1991ax} T.D.~Lee and Y.~Pang, ``Nontopological solitons,''
Phys. Rept. {\bf 221}, 251-350 (1992)

\bibitem{Schunck:2003kk}
F.E.~Schunck and E.W.~Mielke, ``General relativistic boson stars,'' 
Class. Quant. Grav. {\bf 20}, R301-R356 (2003) [arXiv:0801.0307 [astro-ph]]

\bibitem{Liebling:2012fv}
S.L.~Liebling and C.~Palenzuela, ``Dynamical boson stars,'' 
Living Rev. Rel. {\bf 15}, 6 (2012) [arXiv:1202.5809 [gr-qc]]

\bibitem{Torres:2000dw} D.F.~Torres, S.~Capozziello and G.~Lambiase, ``A Supermassive scalar star at the galactic center?,''
Phys. Rev. D {\bf 62}, 104012 (2000) [arXiv:104012 [astro-ph]]

\bibitem{Guzman:2005bs}
F.G.~Guzman, ``Accretion disc onto boson stars: A Way to supplant black holes candidates,''
Phys. Rev. D {\bf 73}, 021501(R) (2006) [arXiv:0512081 [gr-qc]]

\bibitem{AmaroSeoane:2010qx} P.~Amaro-Seoane, J.~Barranco, A.~Bernal, and L.~Rezzolla,
``Constraining scalar fields with stellar kinematics and collisional dark matter,'' JCAP {\bf 1011}, 002 (2010) [arXiv:1009.0019 [astro-ph.CO]]

\bibitem{Hogan:1988mp} C.J.~Hogan and M.J.~Rees, ``Axion miniclusters,''
Phys. Lett. B {\bf 205}, 228-230 (1988)

\bibitem{Kolb:1993zz} E.W.~Kolb and I.I.~Tkachev, ``Axion miniclusters and Bose stars,''
Phys. Rev. Lett. {\bf 71} 3051-3054 (1993) [arXiv:9303313 [hep-ph]]

\bibitem{Barranco:2010ib}
J.~Barranco and A.~Bernal, ``Self-gravitating system made of axions,''
Phys. Rev. D {\bf 83}, 043525 (2011) [arXiv:1001.1769 [astro-ph.CO]]


\bibitem{Sin:1992bg} S-J. Sin, ``Late time cosmological phase transition and galactic halo as Bose liquid,''
Phys. Rev. D {\bf 50}, 3650-3654 (1994) [arXiv:9205208 [hep-ph]]

\bibitem{Schive:2014dra}
H-Y.~Schive, T.~Chiueh and T.~Broadhurst, ``Cosmic structure as the quantum interference of a coherent dark wave,''
Nature Phys. {\bf 10}, 496-499 (2014) [arXiv:1406.6586 [astro-ph.GA]]

\bibitem{Marsh:2015xka} D.J.E.~Marsh, ``Axion cosmology,'' 
Phys. Rept. {\bf 643}, 1-79 (2016) [arXiv:1510.07633 [astro-ph.CO]]

\bibitem{Hui:2016ltb} L.~Hui, J.P.~Ostriker, S.~Tremaine, E.~Witten, ``Ultralight scalars as cosmological dark matter,''
Phys. Rev. D {\bf 95}, 043541 (2017) [arXiv:1610.08297 [astro-ph.CO]]

\bibitem{Olabarrieta:2007di} I.~Olabarrieta, J.F.~Ventrella, M.W.~Choptuik 
and W.G.~Unruh, ``Critical behavior in the gravitational collapse of a scalar 
field with angular momentum in spherical symmetry,''
  Phys.\ Rev.\ D {\bf 76}, 124014 (2007) [arXiv:0708.0513 [gr-qc]] 
  
\bibitem{Derrick:1964ww} G.H.~Derrick, ``Comments on nonlinear wave equations as models for
  elementary particles,'' J. Math. Phys. {\bf 5} 1252-1254 (1964)
  
\bibitem{Diez-Tejedor:2013sza} A.~Diez-Tejedor and A.~Gonzalez-Morales, ``No-go theorem for static scalar field dark matter halos with no Noether charges,'' Phys. Rev. D {\bf 88}, 067302 (2013) [arXiv:1306.4400 [gr-qc]]  

\bibitem{Bizon:2000es}
P.~Bizo\'n and A.~Wasserman, ``On existence of mini-boson stars,''
Commun. Math. Phys. {\bf 215}, 357-373 (2000) [arXiv:0002034 [gr-qc]]

\bibitem{Bernal:2009zy} A.~Bernal, J.~Barranco, D.~Alic and C.~Palenzuela, ``Multi-state boson stars,'' 
Phys. Rev. D {\bf 81} 044031 (2010) [arXiv:0908.2435 [gr-qc]]  

\bibitem{UrenaLopez:2010ur} L.A.~Urena-Lopez and A.~Bernal, ``Bosonic gas as a galactic dark matter halo,''
Phys. Rev. D {\bf 82} 123535 (2010) [arXiv:1008.1231 [gr-qc]]

\bibitem{eCnOoS13}
E.~Chaverra, N.~Ortiz and O.~Sarbach, ``Linear perturbations of self-gravitating spherically symmetric configurations",  Phys.\ Rev.\ D {\bf 87}, 0440154 (2013) [arXiv:1209.3731 [gr-qc]] 



\end{thebibliography}
\end{document}